\documentclass[aip,pop,reprint,10pt,superscriptaddress,showpacs]{revtex4-1}
\usepackage{amsfonts} 
\usepackage{amsmath}
\usepackage{amssymb}
\usepackage{graphicx}
\usepackage{subfigure}
\usepackage{color}
\newcommand*\xbar[1]{%
  \hbox{%
    \vbox{%
      \hrule height 0.5pt 
      \kern0.5ex
      \hbox{%
        \kern-0.1em
        \ensuremath{#1}%
        \kern-0.1em
      }%
    }%
  }%
} 

\begin{document}
\title{Nonlinear  Landau damping  of  wave envelopes   in  a quantum plasma}
\author{Debjani Chatterjee}
\affiliation{Department of Mathematics, Siksha Bhavana, Visva-Bharati University, Santiniketan-731 235, WB, India}
\author{A. P. Misra}
\email{apmisra@visva-bharati.ac.in; apmisra@gmail.com}
\affiliation{Department of Mathematics, Siksha Bhavana, Visva-Bharati University, Santiniketan-731 235, WB, India}
\begin{abstract}
The nonlinear theory of Landau damping of electrostatic  wave envelopes (WEs)  is revisited in a quantum electron-positron (EP) pair plasma. Starting from a   Wigner-Moyal equation coupled to the Poisson equation and applying the multiple scale technique, we derive a nonlinear Schr{\"o}dinger (NLS) equation  which governs the evolution of electrostatic WEs. It is shown that the coefficients of the NLS equation,  including the nonlocal nonlinear term, which appears due to the resonant particles having group velocity of the WEs,  are significantly modified by the particle dispersion. The effects of the quantum parameter $H$ (the ratio of the plasmon energy to the thermal energy densities), associated with the particle dispersion, are examined  on   the Landau damping rate of carrier waves, as well as on the modulational instability  of WEs.  It is found that the Landau damping rate     and  the decay rate  of the solitary  wave amplitude  are  greatly reduced  compared to their classical values $(H=0)$.
\end{abstract}
\maketitle
\section{Introduction} \label{sec-introduct}
Landau damping is one of the most fundamental phenomena of waves in plasma physics. Such collisionless damping was first   theoretically  predicted by Landau \cite{landau1946} and later   experimentally verified  by Malmberg and Wharton \cite{malmberg1964}. Since then, Landau damping of electrostatic or electromagnetic waves in plasmas has been a topic of important research (see, e.g., Refs. \onlinecite{rightley2016,chatterjee2015,villani2014,zheng2013,valentini2007}). However, in most of these investigations Landau damping has been considered classically, i.e., using the Vlasov-Poisson system and/or limited to the linear theories. When quantum effects are included there appear new length scale and new coupling parameter, as well as new collective modes for which new processes come into play. For example,       the modifications of the linear Landau damping of electrostatic waves by the effects of arbitrary degeneracy of electrons    \cite{rightley2016}, and the influence of linear Landau damping on nonlinear waves \cite{mukherjee2014} in quantum plasmas. Although, there are  some    developments of Landau damping in   nonlinear regimes (see, e.g.,  Refs. \onlinecite{chatterjee2015,villani2014,brodin2015}), however, many of the effects of nonlinear Landau damping   have not yet been explored, especially in the quantum regime.  
\par
 Recently,    Chatterjee and Misra \cite{chatterjee2015} advanced the nonlinear theory of Landau damping \cite{ichikawa1974}     of electrostatic wave envelopes in an electron-positron (EP) pair plasma in the context of Tsallis' nonextensive statics. Starting from a set of Vlasov-Poisson equations and using a multiple scale technique, they have shown that the nonextensive parameter significantly modifies the Landau damping rates,   the modulational instability (MI), as well as the evolution of wave envelopes.  

In the present work, we consider the Wigner-Moyal equation,  which accounts for the particle dispersion, and excludes the exchange-correlation and spin effects. Specifically, we focus on the weak quantum regime in which the Langmuir wavelength $(L)$ is larger than the thermal de Broglie wavelength of electrons and positrons $(\lambda_B)$, i.e., $L>\lambda_B\equiv\hbar/mv_t$, where $\hbar$ is the reduced Planck's constant, $m$ is the  mass of electrons/positrons and $v_t=\sqrt{k_BT/m}$ is the thermal velocity with $k_B$ denoting the Boltzmann constant and $T$ the particle's thermodynamic temperature.  Such weak quantum effects in the nonlinear regime can lead to, e.g., bounce-like amplitude oscillations even in absence of trapped particles \cite{brodin2015}.    Starting from the Wigner-Moyal equation coupled to the Poisson equation and using a multiple scale technique (MST), we derive a   nonlinear Schr{\"o}dinger (NLS) equation, which governs the evolution of electrostatic wave envelopes (WEs) in quantum EP plasmas.    We show that in the weak quantum regime, the resonant particles still have a velocity close to the group velocity of the WEs as in the classical theory \cite{chatterjee2015,ichikawa1974}, and the coefficients of the NLS equation, including the nonlocal nonlinearity due to the wave-particle resonance, are significantly modified by the quantum dispersion.  We also find that, not only the wave dispersion and the Landau damping rate are modified, the MI as well as the decay rate of WEs are also greatly influenced by the quantum effect.     
 \section{Basic Equations and derivation of NLS equation}
We consider the nonlinear propagation of electrostatic waves in an unmagnetized collisionless  quantum electron-positron-pair plasma. Such ion-free EP-pair plasmas with unique characteristics have recently been produced in the laboratory\cite{sarri2015}.  Our starting point is the  Wigner-Moyal distribution function $F_{\alpha}(x,v,t)$ for electrons $(\alpha=e)$ and positrons $(\alpha=p)$ which satisfies the following evolution equation 
\begin{equation}
\begin{split}
\frac{\partial F_{\alpha}}{\partial t}+v\frac{\partial F_{\alpha}}{\partial x}+\frac{e_{\alpha}m_{\alpha}}{2i\pi\hbar^2}\int\int dx_0 dv_0 e^{im_{\alpha}(v-v_0)x_0/\hbar}\\  
\times\left[\phi\left(x+\frac{x_0}{2}\right)-\phi\left(x-\frac{x_0}{2}\right)\right]F_{\alpha}(x,v_0, t)=0, \label{wigner-eq-moyal}
\end{split}
\end{equation}
where $e_\alpha$ and $m_\alpha$ are the charge and mass of $\alpha$-species particles, $v$ is the velocity and $\phi(x,t)$ is the self-consistent electrostatic potential satisfying the Poisson equation
\begin{equation}
 {\frac{\partial^2 \phi}{\partial x^2}}=-4\pi\sum e_\alpha \int F_\alpha dv. \label{poisson-eq}
\end{equation} 
 Note that Eq. \eqref{wigner-eq-moyal} includes the particle dispersive effects but ignores the exchange-correlation and spin effects.  We focus on the long-wavelength perturbations with characteristic length scale $L\gg\lambda_D\sim\hbar/mv_t$, where $\lambda_D\equiv v_t/\omega_p=\sqrt{k_BT/8\pi n_0e^2}$ is the typical Debye length with $n_0$ denoting the equilibrium number density, and $\omega_p = \sqrt{8\pi n_0 e^2/m}$    the   oscillation frequency of electrons and positrons.    This assumption requires the smallness of the nondimensional parameter $H=\hbar/mv_t\lambda_D$.
 \par
 In order to obtain the evolution equation for the electrostatic WEs in EP plasmas, we follow  Refs. \onlinecite{ichikawa1974,chatterjee2015} and so, introduce the multiple space-time scales as
 \begin{equation}
  x\rightarrow x+\epsilon ^{-1} \eta +\epsilon^{-2}\zeta,~ ~t\rightarrow t+\epsilon ^{-1}\sigma,    \label{stretch}
 \end{equation}
where $\eta$, $\zeta$, and $\sigma$ are coordinates stretched by a small parameter $\epsilon$.
Since the wave amplitude is infinitesimally small so for $t>0$   a slight deviation  $\sim o(\epsilon)$ from the uniform initial sate will occur. Thus, we expand   
\begin{equation}  
\begin{split}
&F_\alpha(v,x,t)= F^{(0)}_\alpha (v)+\sum_{n=1}^{\infty}\epsilon^{n} \sum_{l=-\infty}^{\infty}f^{(n)}_{\alpha,l}(v,\eta,\sigma,\zeta)\\
&\times \exp[il(kx-\omega t)], \\ 
&\phi(x,t)=\sum_{n=1}^{\infty}\epsilon^{n} \sum_{l=-\infty}^{\infty}\phi^{(n)}_l(\eta,\sigma,\zeta) \exp[il(kx-\omega t)], \label{expansion} 
\end{split}
\end{equation}
where $\omega~(k)$ is the carrier wave frequency (number) of perturbations and  $f^{(n)}_{\alpha,-l}=f^{(n)\ast}_{\alpha,l}$, $\phi^{(n)}_{-l}=\phi^{(n)\ast}_l$ are the reality conditions to hold.  Note that the expansion \eqref{expansion} and the stretched coordinates  $\xi=\epsilon(x-\lambda t),~\tau=\epsilon^2 t$ (not considered here) are usually used in the reductive perturbation technique for the derivation of, e.g., the  NLS equation. However, it has been shown by Ichikawa {\it et al.} \cite{ichikawa1972}  that the direct application of the reductive perturbation technique with the expansion \eqref{expansion} to the Vlasov-Poisson system does not determine uniquely the contributions of resonant particles having group velocity of the wave  $(v\sim\lambda)$. In order to give a proper account of this resonance,  Ichikawa  \cite{ichikawa1974} proposed a new multiple space-time scale \eqref{stretch} with  further  expansions of $f^{(n)}_{\alpha,l}$ and $\phi^{(n)}_l$ in terms of Fourier-Laplace integrals.   
 So, the components $f^{(n)}_{\alpha,l}$ and $\phi^{(n)}_l$ are further expanded  as
  \begin{equation}
 \begin{split}
 f^{(n)}_{\alpha,l}(v,\eta,\sigma,\zeta)=& \frac {1}{(2\pi)^2} \int_C d\Omega \int_ {-\infty}^{\infty}dK\tilde{f}^{(n)}_{\alpha,l}(v,K,\Omega,\zeta)\\
 &\times \exp[i(K\eta-\Omega\sigma)] \\
 \phi^{(n)}_l(\eta,\sigma,\zeta)= &\frac {1}{(2\pi)^2} \int_C d\Omega \int_{-\infty}^{\infty}dK\tilde{\phi}^{(n)}_l (K,\Omega,\zeta)\\
 &\times \exp[i(K\eta-\Omega\sigma)], \label{Fourier-Lap-int}
\end{split}
 \end{equation}
where the contour $C$ is taken parallel to the real axis lying above the coordinate of convergence.
\par
 In the weak quantum regime in which $\hbar k/mv_t<1$ (Note here that in the true quantum regime  where the thermal de Broglie wavelength is comparable to or larger than the typical Langmuir wavelength of EP plasma oscillations, i.e., $\hbar k/mv_t\gtrsim1$, the Wigner-Moyal equation  \eqref{wigner-eq-moyal} is retained in its full form.), the integrand  in Eq. \eqref{wigner-eq-moyal} can be Taylor expanded to retain terms up to $o(H^2)$. Thus, from Eq. \eqref{wigner-eq-moyal} we obtain 
 \begin{equation}
\frac{\partial F_{\alpha}}{\partial t}+v\frac{\partial F_{\alpha}}{\partial x}-\frac{e_\alpha}{m_\alpha}\frac{\partial \phi}{\partial x} \frac{\partial F_{\alpha}}{\partial v}+\frac{e_\alpha\hbar^2 }{24 m_\alpha^3} \frac{\partial^3 \phi}{\partial x^3} \frac{\partial^3 F_{\alpha}}{\partial v^3} +o(H^4)=0,  \label{wigner-eq}
\end{equation}
from which the Vlasov equation can be recovered  in the formal semiclassical limit $\hbar\rightarrow0$.
\par 
  Substituting the stretched coordinates \eqref{stretch} and   the expansions \eqref{expansion} into Eqs. \eqref{poisson-eq} and \eqref{wigner-eq}     we obtain, respectively, 
\begin{equation}
\begin{split}
&il(\omega-kv)f^{(n)}_{\alpha,l}+ilkG_\alpha \phi^{(n)}_l-il^3k^3H_\alpha\phi^{(n)}_l\\
&\doteq\frac{\partial }{\partial \sigma}f^{(n-1)}_{\alpha,l}+v\frac{\partial }{\partial \eta}f^{(n-1)}_{\alpha,l}+v\frac{\partial }{\partial \zeta}f^{(n-2)}_{\alpha,l}-G_\alpha\frac{\partial }{\partial \eta}\phi^{(n-1)}_l\\
&-G_\alpha\frac{\partial }{\partial \zeta}\phi^{(n-2)}_l 
-ik\frac{e_\alpha}{m_\alpha}\sum^\infty_{s=1}\sum^\infty_{l'=-\infty}(l-l')\phi^{(n-s)}_{l-l'} \frac{\partial }{\partial v}f^{(s)}_{\alpha,l'}\\
&-\frac{e_\alpha}{m_\alpha}\sum^\infty_{s=1}\sum^\infty_{l'=-\infty}\left(\frac{\partial}{\partial \eta} \phi^{(n-s-1)}_{l-l'}+ \frac{\partial }{\partial \zeta}\phi^{n-s-2}_{l-l'}\right)  \frac{\partial }{\partial v}f^{(s)}_{\alpha,l'}\\
&+3lkH_\alpha\left(lk \frac{\partial}{\partial \eta}\phi^{(n-1)}_l +lk \frac{\partial}{\partial \zeta}\phi^{(n-2)}_l-i \frac{\partial^2}{\partial \eta^2}\phi^{(n-2)}_l\right)\\
&-i\frac{e_\alpha\hbar^2}{24 m_\alpha^3}k^3\sum^\infty_{s=1}\sum^\infty_{l'=-\infty}{(l-l')^3}\phi^{(n-s)}_{l-l'} \frac{\partial^3}{\partial v^3}f^{(s)}_{\alpha, l'}\\
&-3\frac{e_\alpha\hbar^2}{24 m_\alpha^3}k^2\sum^\infty_{s=1}\sum^\infty_{l'=-\infty}{(l-l')^2}\frac{\partial}{\partial \eta}\phi^{(n-s-1)}_{l-l'} \frac{\partial^3}{\partial v^3}f^{(s)}_{\alpha, l'}\\
&-3\frac{e_\alpha\hbar^2}{24 m_\alpha^3}k^2\sum^\infty_{s=1}\sum^\infty_{l'=-\infty}{(l-l')^2}\frac{\partial}{\partial \zeta}\phi^{(n-s-2)}_{l-l'} \frac{\partial^3}{\partial v^3}f^{(s)}_{\alpha, l'}\\
&+i3k\frac{e_\alpha\hbar^2}{24 m_\alpha^3}\sum^\infty_{s=1}\sum^\infty_{l'=-\infty}{(l-l')}\frac{\partial^2}{\partial \eta^2}\phi^{(n-s-2)}_{l-l'} \frac{\partial^3}{\partial v^3}f^{(s)}_{\alpha, l'}, \label{vlasov1}
\end{split}
\end{equation}
 \begin{equation}
 \begin{split}
&(lk)^2\phi^{(n)}_l-2ilk\frac{\partial}{\partial \eta}\phi^{(n-1)}_l -i2lk \frac{\partial }{\partial \zeta}\phi^{(n-2)}_l \\
&-\frac{\partial^2}{\partial \eta^2}\phi^{(n-2)}_l
-4\pi\sum_{\alpha} e_\alpha\int f^{(n)}_{\alpha,l}dv=0,  \label{poisson1}
 \end{split}
\end{equation}
where 
\begin{equation}
G_\alpha (v)=\frac{e_\alpha}{m_\alpha}\frac {\partial}{\partial v}F^{(0)}_\alpha (v),~~H_\alpha (v)=-\frac{e_\alpha\hbar^2}{24 m^3_\alpha}\frac {\partial^3}{\partial v^3}F^{(0)}_\alpha (v).\label{GH-alpha}
\end{equation}
The symbol $\doteq$ in Eq. \eqref{vlasov1} is used to denote the equality in the weak sense, and in Eq. \eqref{poisson1} we have disregarded the terms  containing $\phi^{(n-3)}_l$ and $\phi^{(n-4)}_l$.   
 
In what follows, we determine the contributions of the resonant particles having the group velocity of the wave envelopes by solving the $\sigma$-evolution of the components   $f^{(n)}_{\alpha,l}$ and $\phi^{(n)}_l$ as an initial value problem  with the  initial condition
\begin{equation}
f^{(n)}_{\alpha,0} (v, \eta, \sigma=0, \zeta)\doteq0,~~~ n\geq 1, \label{int-cond1}
\end{equation}
in the multiple space-time scheme corresponding to that on the distribution function
\begin{equation}
f^{(n)}_{\alpha,0} (v, t=0)=0. \label{int-cond2}
\end{equation}
In the following subsections, we will skip the    detailed derivations of the expressions for different kinds of modes and relevant others. For some details, readers are referred to, e.g.,   Refs. \onlinecite{ichikawa1974,chatterjee2015}.
\subsection{Harmonic modes with $n=1,~l=1$: Linear dispersion law}
From Eqs. \eqref{vlasov1} and \eqref{poisson1}   equating the coefficients of $\epsilon$ for $n=1,~l=1$,  we obtain the linear dispersion law  

\begin{equation}
 k+4\pi\sum e_\alpha\int_C\frac{G_\alpha(v)-k^2 H_\alpha(v)}{\omega-kv}dv=0. \label{dispersion}
\end{equation}
with the Landau damping rate $\gamma_L\sim\omega_r\epsilon^{2+p}$, where $\omega_r=\Re{\omega}$ and $p$ is a non-negative integer, given by \cite{chatterjee2015}
\begin{equation}
\begin{split}
 \gamma_L=&\frac{\pi}{k}\sum_{\alpha}e_{\alpha}\left\lbrace G_{\alpha}\left(\frac{\omega_r}{k}\right)-k^2H_\alpha\left( \frac{\omega_r}{k}\right) \right\rbrace\\
 & \Big/\sum_{\alpha}e_{\alpha}\int_C\frac{G_{\alpha}-k^2H_\alpha}{\left(\omega_r-kv\right)^2}dv,\label{dispersion-img}
 \end{split}
 \end{equation}
where  $F^{(0)}_\alpha (v)$  is the unperturbed part of the distribution function $F_\alpha (v)$ and  $\int_C$ denotes the Cauchy Principal value along the real axis passing infinitesimally above and under the pole at $v=\omega_r/k$ with the constraint of weakly damped waves. 
\par
Next, considering the harmonic modes for $l\neq 0,~n=1$ and the zeroth harmonic modes for $n=1,~2;~l=0$
we obtain the following conditions:
  \begin{equation}
 f^{(1)}_{\alpha,l}\doteq 0~\text{and}~\phi^{(1)}_l=0~\text{for}~|l|\geq 2. \label{cond-f-phi-l-1}
 \end{equation}
together with the zeroth-order components
 \begin{equation}
f^{(1)}_{\alpha,0}\doteq 0,~~~\phi^{(1)}_0=0.  \label{f-phi-n1-l0}
\end{equation}

\subsection{Modes with $n=2,~l=1$: Compatibility condition}
For the second-order, first harmonic modes with $n=2,~l=1$ we obtain    from Eqs. \eqref{vlasov1} and   \eqref{poisson1} the compatibility condition for the group velocity,
\begin{equation}
\left\{ \frac{\partial}{\partial \sigma} + \lambda \frac{\partial}{\partial \eta}\right\} \phi^{(1)}_1(\eta, \sigma;\zeta) =0, \label{vg-eq}
\end{equation}
where $\lambda$ is  given by 
\begin{equation}
\begin{split}
&\lambda\equiv \frac{\partial\omega_r}{\partial k}=\left[4\pi\sum e_\alpha \int_C\frac{\left(G_\alpha-k^2H_\alpha\right)}{\left(\omega_r-kv\right)^2} dv\right]^{-1}\\ &\times\left[1+4\pi\sum e_\alpha \int_C \frac{vG_\alpha-kH_\alpha\left(2\omega_r-kv\right)}{\left(\omega_r-kv\right)^2}dv\right]. \label{lambda}
\end{split}
\end{equation}
From Eq. \eqref{vg-eq} it is clear that    the $\sigma-\eta$ variation of $\phi^{(1)}_1$  can be related  to a new coordinate $\xi=\eta-\lambda \sigma= \epsilon(x-\lambda t)$ such that $
\phi^{(1)}_1 (\eta, \sigma;\zeta)=\phi^{(1)}_1 (\xi;\zeta)$. Thus, the  coordinate $\xi$   establishes a clear relationship between the reductive perturbation theory and the multiple space-time expansion method.
\subsection{Second  harmonic modes with $n=l=2$}
The second-order perturbed quantities with $n=l=2$ can be obtained from  Eqs. \eqref{vlasov1} and \eqref{poisson1} as
\begin{equation}
\begin{split}
 f^{(2)}_{\alpha,2}\doteq & - \frac{k}{\omega-kv+i\nu}\left[(G_\alpha-4k^2H_\alpha) \phi^{(2)}_2 \right.\\
 & \left.-\frac{ke_\alpha}{2 m_\alpha} \frac{\partial}{\partial v}\left\lbrace  \frac{G_\alpha-k^2H_\alpha}{\omega-kv+i\nu}\right.\right.\\
 &\left.\left.+\frac{k \hbar^2}{24m_\alpha^2}\frac{\partial^2}{\partial v^2}\left(\frac{G_\alpha-k^2 H_\alpha}{\omega-kv+i\nu} \right)\right\rbrace   \left(\phi^{(1)}_1\right)^2\right], \label{f-alpha2-n2-l2}
 \end{split}
\end{equation}
\begin{equation}
\phi^{(2)}_2=\frac{1}{6} A(k,\omega)\left(\phi^{(1)}_1\right)^2, \label{phi2-n2-l2}
\end{equation}
where
\begin{equation}
\begin{split}
A(k,\omega)=&4\pi\sum_\alpha \frac{e_\alpha ^2}{m_\alpha}\left[\int_C \frac{1}{\omega-kv}\frac{\partial}{\partial v}\left\lbrace\frac{G_\alpha-k^2H_\alpha}{\omega-kv}\right.\right.\\
&\left.\left.+\frac{k^2\hbar^2}{24m_\alpha^2}\frac{\partial^2}{\partial v^2}\left( \frac{G_\alpha-k^2H_\alpha} {\omega-kv}\right)\right\rbrace dv\right]\\
&\Bigg/\left(1-4\pi\sum_\alpha e_\alpha\int_C\frac{kH_\alpha}{\omega-kv}dv\right), \label{A-k-omega}
\end{split}
\end{equation}
where $\nu=\pm|\nu|$ for $l\lessgtr0$ is introduced to anticipate that the solution in the linear approximation decays with the Landau damping rate \cite{chatterjee2015}.
Since the linear Landau damping $\gamma_L$ is assumed to be a higher-order effect than second order,  we have neglected  the effect of the resonance terms at the phase velocity $\omega/k$ in Eq. \eqref{A-k-omega}. 
\subsection{Modes with $n=3,~l=0$}
In what follows, we consider the  harmonic modes with $n=3$ and $l=0$ from Eqs. \eqref{vlasov1} and    \eqref{poisson1}. Then  using the relations \eqref{f-phi-n1-l0} and \eqref{vg-eq} we obtain a set of reduced equations, which after use of the Fourier-Laplace transforms with respect to $\eta$ and $\sigma$ and the initial condition \eqref{int-cond1}, yield
\begin{equation}
\begin{split}
&\tilde{f}^{(2)}_{\alpha,0}\doteq k^2\left[-\frac{{\cal{W}}(K,\Omega)}{\Delta^{(c)}(K,\Omega)} \frac{K}{\Omega-Kv}G_\alpha(v)\right.\\
&\left.-\frac{e_\alpha}{m_\alpha}\frac{K}{\Omega-Kv}I_\alpha(v)\right] \Xi(K,\Omega), \label{fF-alpha-n2-l0-final}
\end{split}  
\end{equation}
\begin{equation}
\tilde{\phi}^{(2)}_0=k^2\frac{\Xi(K,\Omega)}{\Delta^{(c)}(K,\Omega)}{\cal{W}} (K,\Omega), 
\end{equation}
where
\begin{equation}
\begin{split}
 I_\alpha(v)=&\frac{\partial}{\partial v}\left\{\frac{(v-\lambda)\left(G_\alpha-k^2H_\alpha\right)}{(\omega-kv)^2} -\frac{2kH_\alpha}{\omega-kv}\right\} \\
 &+\frac{k^2\hbar^2}{24 m_\alpha^2}\frac{\partial^3}{\partial v^3}\left[ \frac{(v-\lambda)\left(G_\alpha-k^2H_\alpha\right)}{(\omega-kv)^2}\right.\\
 &\left.-\frac{2k}{\omega-kv}\left(G_\alpha+(1-k^2)H_\alpha\right)\right]. \label{I-alpha-v}
 \end{split}
\end{equation}
and  $\Xi(K,\Omega)$ is defined as
\begin{equation}
\begin{split}
 \mid\phi^{(1)}_1(\eta-\lambda \sigma,\zeta) \mid^2=&\frac{1}{(2\pi)^2}\int d\Omega \int dK\Xi(K,\Omega)\\
 &\times\exp[i(K\eta-\Omega\sigma)], \label{phi-1-1-mod}
 \end{split}
\end{equation}
with
\begin{equation}
\Xi(K,\Omega)=2\pi\delta(\Omega-K\lambda)\int dK' \phi^{(1)\ast}_1(K')\phi^{(1)}_1(K+K'), \label{H-K-Omega}
\end{equation}
 \begin{equation}
 \begin{split}
&{\cal{W}}(K,\Omega)=4\pi \sum_\alpha \frac{e^2_\alpha}{m_\alpha} \int \frac{K}{\Omega-Kv}I_\alpha dv,\\
&\Delta^{(c)}(K,\Omega)=-4\pi K\sum_\alpha e_\alpha \int \frac{G_\alpha}{\Omega-Kv}dv. \label{W-cal}
\end{split}
\end{equation}
\subsection{Harmonic modes with $n=3,~l=1$ and the NLS equation} \label{section-nls-equation}
Finally, considering the terms  for $n=3$ and $l=1$ from Eqs. \eqref{vlasov1} and \eqref{poisson1}, and adopting the same procedure as in Ref. \onlinecite{chatterjee2015} we obtain (after rescaling with $\zeta=\lambda\tau$) the following quantum modified NLS equation for the small but finite  amplitude perturbation $\phi(\xi, \tau)\equiv\phi^{(1)}_1(\xi, \tau)$: 
\begin{equation}
i\frac{\partial\phi}{\partial \tau}+P \frac{\partial^2\phi}{\partial \xi^2}+Q |\phi|^2 \phi +\frac{R }{\pi}{\cal P}\int\frac{ |\phi(\xi',\tau)|^2}{\xi-\xi'} d\xi' \phi+iS\phi=0. \label{nls}
\end{equation}
 The coefficients of the group velocity dispersion $(P)$,   local cubic nonlinear $(Q)$ and nonlocal nonlinear $(R)$ terms are  given by $P\equiv(1/2)\partial^2\omega/\partial k^2=\beta/ \alpha,~Q= \gamma/ \alpha$ and $R=\delta/ \alpha$, where
\begin{equation}
\alpha=4\pi \sum_\alpha e_\alpha \int \frac{G_\alpha-k^2H_\alpha}{(\omega-kv)^2}dv, \label{alpha}
\end{equation}
\begin{equation}
\begin{split}
\beta=&4\pi \sum_\alpha e_\alpha \int \left[\frac{(v-\lambda)^2\left(G_\alpha-k^2H_\alpha\right)}{\left(\omega-kv\right)^3}\right.\\
&\left. -\frac{\left(1+2k(v-\lambda)\right)H_\alpha}{\left(\omega-kv\right)^2}\right] dv, \label{beta}
\end{split}
\end{equation}
\begin{equation}
\gamma=\left(\frac{1}{6}\frac{A A_1}{k}+\frac{1}{2}B\right)k^2-\Theta(k,~ \omega),~~\delta=-\Phi(k, \omega). \label{gamma}
\end{equation}
The expressions for   $A_1,~B,~\Theta,~\Phi$ etc. are given in Appendix \ref{appendix-a}.
 Furthermore, the damping coefficient $S$ in Eq. \eqref{nls}   associated with the resonant particles having the phase velocity of the carrier wave, is given by
\begin{equation}
S=\frac{\theta (p)\gamma_L}{\epsilon^2}. \label{S}
\end{equation}
We note that all the coefficients of the NLS equation \eqref{nls} are modified by the quantum parameter $H$ associated with the particle dispersion. Furthermore, the nonlocal term $\propto R$ appears due to the wave-particle resonance having the group velocity of the wave envelopes. This resonance contribution also modifies  the local nonlinear coefficient $Q$, which appears due to the carrier wave self-interactions. 
\section{Electrostatic envelopes with Maxwellian stationary state} \label{sec-langmuir-envelopes}
 We consider the nonlinear propagation of high-frequency $(\omega>\omega_p)$ Langmuir waves whose phase velocity greatly exceeds  the  thermal velocities of electrons and positrons, i.e.,  ${\omega_r}/{k}\gg v_{t}>v$. For analytical simplicity, we also consider the  long-wavelength perturbations, i.e., $k\ll k_d$, where $k_d=\left(8\pi n_0 e^2/k_B T\right)^{1/2}$ is the plasma Debye wave number, and the quantum contribution to the wave-particle interaction is small, i.e., $H^2\lesssim1$. Furthermore, we  assume that the particle's thermodynamic temperature is higher than the Fermi temperature, i.e., $T> T_F$, so that    the equilibrium distributions of electrons and positrons can be considered as the   Maxwellian-Boltzmann, given by, 
\begin{equation}
 F^{(0)}_\alpha (v)= \frac{n_0}{\sqrt{2\pi}v_t} \exp\left( -\frac{v^2}{2v_t^2} \right). \label{maxwellian-distribution}
 \end{equation}
 where for  a fully symmetric and charge-neutral  EP plasma, $T_e=T_p=T$ and $m_e=m_p=m$.   Under these assumptions, the linear dispersion relation, the Landau damping rate and the coefficients of the NLs equation \eqref{nls} can be simplified. For detailed analysis, readers are suggested to Ref. \onlinecite{chatterjee2015}. Thus, in the region of small wave numbers with $\chi^2\equiv k^2/k_d^2\ll1$  and  the smallness of thermal corrections, the dispersion    relation for Langmuir waves \eqref{dispersion} reduces to
\begin{equation}
{\omega_r}^2 = {\omega_p}^2\left( 1+ 3\chi^2 +\frac{1}{4}H^2 \chi^4\right). \label{dispersion-reduced}
\end{equation}
Equation \eqref{dispersion-reduced} is the known form of  the dispersion relation for Langmuir waves in a quantum plasma with particles in thermodynamic equilibrium \cite{manfredi2001}. In the formal limit $H\rightarrow0$, one can recover the dispersion relation for high-frequency Langmuir waves in classical plasmas. However, the Landau damping rate [Eq.\eqref{dispersion-img}], which is one of our main results in quantum EP plasmas, takes the form
\begin{equation}
\begin{split}
\gamma_L=&- \sqrt{\frac{\pi}{8}} \frac{\omega_p}{\chi^3} \exp \left[ {-\frac{1}{2\chi^2}} \left(1+3 \chi^2 +\frac14H^2\chi^4 \right) \right]\\
&\times \left[1+\frac{H^2}{24}- \frac{H^2 \chi^2}{2}\left(\chi^2+\frac14\right)   \right]. \label{landau-damping-reduced}
\end{split}
\end{equation}
 We note that the Landau damping rate is also significantly modified by the quantum correction associated with the particle dispersion. In the limit $H\rightarrow0$, and after a small adjustment of the factor $8$ in $\omega_p$ and under the square root in the expression of $\gamma_L$,  Eq. \eqref{landau-damping-reduced} exactly agrees   with that for Langmuir waves in classical plasmas \cite{chen1984}.   
 \par
\begin{figure*}[ht]
\centering
\includegraphics[height=2.2in,width=6.5in]{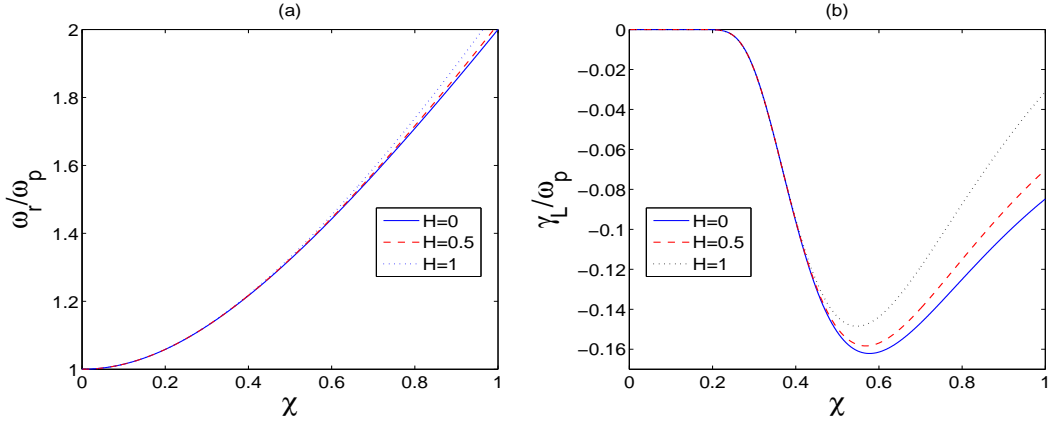}
\caption{ (Color online) The real [$\omega_r/\omega_p$, Eq. \eqref{dispersion-reduced}, panel (a)] and imaginary [$\gamma_L/\omega_p$, the linear Landau damping rate given by Eq. \eqref{landau-damping-reduced}, panel (b)] parts of the carrier wave frequency $\omega$ are plotted against $\chi\equiv k/k_d$ for different values of $H$ as in the legends. }
\label{fig1}
\end{figure*}

  We numerically analyze the dispersion properties and the Landau damping rate of Langmuir waves  by the influence of the quantum parameter $H$.   The profiles of the dispersion relation [Eq. \eqref{dispersion-reduced}] and the Landau damping rate [Eq. \eqref{landau-damping-reduced}] are shown in Fig.  \ref{fig1} for different values of $H$ that correspond to different plasma environments as represented by the plasma number density $n_0$ and the  temperature $T$. For example, $H=0.5$ corresponds to the regime where $T=10^6$ K, $T_F/T=0.3$ and $ n_0=7\times10^{23}$ cm$^{-3}$, and $H=1$  represents one with $T=6\times10^5$ K, $T_F/T=0.7$ and $ n_0=10^{24}$ cm$^{-3}$. For a fixed $T~(n_0)$, as $n_0~(T)$ increases (decreases), the values of the ratios $H$ and $T_F/T$ increase. It is found that both the real part of the wave frequency and  the absolute value of the damping rate decreases with increasing values of $H$ in $0\lesssim H\lesssim1$.    In fact, there are two subregions of $\chi$, namely $0\lesssim\chi\lesssim0.6$ and $0.6<\chi\lesssim1$. In the former,   the value of $|\gamma_L|$ increases, whereas the same decreases in the other one  where    the effects of $H$ become prominent.   Thus, we conclude that for a fixed temperature $T$, as one enters the high-density regime or with a fixed density, as one look for low-temperature plasmas, the quantum contribution to the wave-particle interaction is that it influences the wave to lose energy to the particles more slower than the classical theory.  
\par
In what follows, we simplify the  coefficients of the NLS equation \eqref{nls} in the limit $\chi^2\ll1$ and $H^2\lesssim1$. From the expression of   $P\equiv\beta/\alpha$  we obtain (for some details, see Appendix \ref{appendix-b})
\begin{equation}
P=\frac{3}{2}\frac{\omega_p}{k_d^2} \left[1- \frac{1}{2}(9-H^2) \chi^2+\frac{85}{8} H^2 \chi^4\right]. \label{P-reduced}
\end{equation}
For the other coefficients, namely $Q$ and $R$, we first  obtain the expressions for $\Theta(k,\omega)$ and $\Phi(k,\omega)$ in the limit $\chi^2\ll1$. So, we calculate, respectively, the resonant and non-resonant contributions to $\Theta(k,\omega)$ as
\begin{equation}
\frac{k^3}{\Delta^2+\Gamma^2}\left(\Gamma WW_3-\Delta UU_3\right)^2=0, \label{reso-contri}
\end{equation}
\begin{equation}
k^3\left[\frac{(WW_3)^2}{\Delta}+C_3\right]=\left(\frac{e}{k_BT}\right) ^2\chi^2k \left[1-\frac{H^2}{4} \chi^2  \right]. 
\end{equation}
Thus, it follows that the resonant contribution to  the coefficient  $Q$, which is smaller than that of the nonresonant one,    can be disregarded to obtain
\begin{equation}
Q= -\frac{1}{2} \left( \frac{e}{k_BT}\right) ^2 \omega_p \chi^2 \left( 1-\frac{H^2}{4} \chi^2\right). \label{Q-reduced}
\end{equation}
The contribution from the group velocity resonance through $\Phi(k,\omega)$ gives 
\begin{equation}
R=\frac{3}{2} \left( \frac{e}{k_BT}\right) ^2 \left( \frac{\pi}{2}\right) ^{1/2} \omega_p\chi^3 \left( 1-\frac{13H^2}{24}\chi^2\right). \label{R-reduced} 
\end{equation}
From Eqs. \eqref{P-reduced}, \eqref{Q-reduced} and \eqref{R-reduced}, it is clear that $Q<0$ and $R>0$ for any values of $\chi$ and $H$ in the interval $(0~1]$. However, $P$ can be either positive or negative depending on the values of $H$ and $\chi$. Figure \ref{fig2} shows the regions for $PQ>0$, i.e., $P>0$ and $PQ<0$, i.e., $Q<0$ in the $\chi H$ plane. In Sec. \ref{sec-MI}, we find that, though the condition for the MI is independent on the sign of $PQ$, but on the presence of $R$, the sign of $PQ$ can be important for determining the natures of the frequency shift $(\Omega_r)$ and the energy transfer rate $(\Gamma)$ of modulated waves. Furthermore, the sign of $R$ is also important to determine whether a steady state solution of the NLS equation exists or not. It has been shown in Ref. \onlinecite{chatterjee2015} that the condition for  an initial perturbation to decay  with time requires $R>0$. From Fig. \ref{fig2}, it is seen that   $P<0$  in a wide range of values of $\chi$ and $H$, i.e., roughly in $0\lesssim\chi\lesssim0.47$, $0\leq H\leq1$ and $0.47\lesssim\chi\lesssim1$, $0.7\leq H\leq1$, otherwise $P>0$, i.e., roughly in    $0.47\lesssim\chi\lesssim1$, $0\leq H<0.7$. Thus, the quantum parameter shifts the positive and negative regions of $PQ$ around the values of $\chi$.

It is instructive to verify the conservation laws, namely the mass, the momentum and the energy that are associated with the NLS equation \eqref{nls}. It can be shown that the nonlocal nonlinear term $\propto R$   violates the energy conservation law  \cite{chatterjee2015}, i.e., the time derivative of the  integral $\partial I_3/\partial\tau<0$ for $R>0$, where $I_3=\int\left[|\partial_\xi\phi|^2-\left(Q/2P\right)|\phi|^4\right]d\xi$. This implies that  an initial perturbation (e.g., in the form of a plane wave) will decay to zero with time, and hence a steady state solution of the NLS equation \eqref{nls} with $|I_3|<\infty$ may not be possible.   
\section{The nonlinear landau damping and modulational instability} \label{sec-MI}
We consider the amplitude modulation of electrostatic wave envelopes in EP plasmas. The relevant details is available in the literature \cite{ichikawa1974,chatterjee2015}. Here, we assume that the linear Landau damping rate is higher order than $\epsilon^2$ and  a plane wave solution of  Eq. \eqref{nls}  is  of the   form  
\begin{equation}
\phi= \rho^{1/2}\exp\left(i \int ^\xi \frac{\sigma}{2P} d\xi\right), \label{sol-nls}
\end{equation}
where $\rho$ and $\sigma$ are real functions of $\xi$ and $\tau$.  Next, we substitute the solution \eqref{sol-nls} into Eq. \eqref{nls} to obtain a set of equations for the real and imaginary parts. Linearizing this set of equations by   splitting up $\rho$ and $\sigma$ into their equilibrium (with suffix $0$) and perturbation (with suffix $1$) parts, i.e., 
\begin{equation}
\rho= \rho_0 +\rho_1 \cos{(K\xi-\Omega \tau)}+\rho_2\sin{(K\xi-\Omega \tau)}, \label{rho-perturbation}
\end{equation}
\begin{equation}
\sigma= \sigma_1 \cos{(K\xi-\Omega \tau)}+\sigma_2 \sin{(K\xi-\Omega \tau)}, \label{sigma-perturbation}
\end{equation}
where $\Omega$ and $K$ are, respectively, the wave frequency and the  wave number of modulation, we obtain, after few steps, the following  dispersion relation for modulated wave envelopes in quantum EP plasmas \cite{ichikawa1974,chatterjee2015}
\begin{equation}
(\Omega^2 +2\rho_0PQK^2-P^2K^4)^2=-(2\rho_0PR K^2)^2. \label{dispersion-nls}
\end{equation}
From Eq. \eqref{dispersion-nls}  it is evident that, due to the nonzero coefficient $R$ associated with the resonant particles having the wave group velocity, the Langmuir wave packet is always unstable irrespective of the signs of    $P$ and $Q$.  A general solution of  Eq. \eqref{dispersion-nls}  can be obtained by considering $\Omega= \Omega_r + i\Gamma $, with $\Omega_r,~\Gamma$ being real,  as
\begin{equation}
\begin{split}
\Omega_r=&\frac{1}{\sqrt{2}}\left[\left\lbrace \left(P^2 K^2-2\rho_0PQ\right)^2+\left(2\rho_0PR\right)^2\right\rbrace^{1/2}\right.\\
&\left. +\left(P^2 K^2-2\rho_0PQ\right)\right]^{1/2}K, \label{Omega-real}
\end{split}
\end{equation}
\begin{equation}
\begin{split}
\Gamma=&-\frac{1}{\sqrt{2}}\left[\left\lbrace\left(P^2 K^2-2\rho_0PQ\right)^2+\left(2\rho_0PR\right)^2\right\rbrace^{1/2}\right.\\
&\left.-\left(P^2 K^2-2\rho_0PQ\right)\right]^{1/2}K. \label{Omega-img}
\end{split}
\end{equation}
\par
Some particular cases may be of interest. In the  small amplitude limit with $\rho_0\ll |P/2Q|K^2$, Eqs. \eqref{Omega-real} and \eqref{Omega-img} reduce to  
\begin{equation}
\Omega_r\approx\  |P|K^2,~\Gamma\approx-\rho_0R. \label{Omega-small-amp}
\end{equation}
In this case, the frequency shift (the real part of $\Omega$) is related to the group velocity dispersion, while the imaginary part describes the nonlinear Landau damping process in which the wave energy is transferred from the higher frequency sidebands to lower frequency ones.   Since $P$  turns  over from negative to positive   values in the  $\chi H$ plane, the values of $\Omega_r$ can be  reduced with cutoffs in one subregion, while in the other its values may be increased with $\chi$. Furthermore, in the region of $\chi$ where $P<0~(P>0)$, the frequency shift increases (decreases) with increasing values of the quantum parameter $H$.   Also, since $R$ increases with $\chi$ but decreases with increasing values of $H$, the   energy transfer rate can be slower by the quantum dispersive effect. 
\par
\begin{figure}[ht]
\centering
\includegraphics[height=2.0in,width=3.5in]{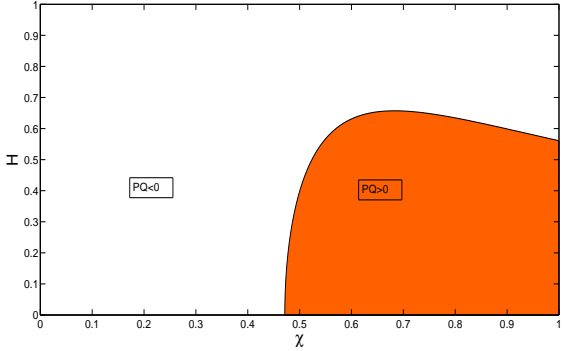}
\caption{ (Color online) Contour plot of $PQ=0$ to show the regions for $PQ>0$ (shaded or colored) and  $PQ<0$ (blank  or white) in the $\chi-H$ plane.      }
\label{fig2}
\end{figure}

In the large amplitude limit with $\rho_0\gg |P/2Q|K^2$ and for $PQ<0$,  we obtain
\begin{equation}
\Omega_r=\sqrt{\rho_0(-PQ)K^2} \left[\sqrt{1+\left({R}/{Q}\right)^2}+1\right]^{1/2}, \label{Omega-real-final}
\end{equation}
\begin{equation}
\Gamma= -\sqrt{\rho_0(-PQ)K^2} \left[\sqrt{1+\left({R}/{Q}\right)^2}-1\right]^{1/2}. \label{Omega-img-final}
\end{equation}
Here, the values of $\Omega_r$ and $\Gamma$ exist in the regions of $\chi$ and $H$ where $PQ<0$, i.e., roughly in $0\lesssim\chi\lesssim0.47$, $0\lesssim H\lesssim1$  and  $0.47\lesssim\chi\lesssim1$, $0.7\lesssim H\lesssim1$ (see the white or blank region in  Fig. \ref{fig2}).
\par
  On the other hand, for $PQ>0$ and for a given value of $\rho_0$, the maximum values of  $\Omega_r$ and $\Gamma$ can be obtained from Eqs. \eqref{Omega-real} and \eqref{Omega-img} at $K=K_m$  as
\begin{equation}
\Omega_m=\pm \frac{R}{Q}\left(Q^2+R^2\right)^{1/2}\rho_0, \label{Omega-max}
\end{equation}
\begin{equation}
\Gamma_m=\mp\left(Q^2+R^2\right)^{1/2}\rho_0, \label{Gamma-max}
\end{equation}
where $K_m$ is given by \begin{equation}
{K_m}^2= \rho_0\left(\frac{Q^2+R^2}{PQ}\right). \label{K-max}
\end{equation}
Thus, maximum values of both $\Omega_r$ and $\Gamma$ can be obtained in some regions of $\chi$ and $H$, e.g.,  $0.47\lesssim\chi\lesssim1$ and $0\lesssim H\lesssim0.7$ where $PQ>0$ (see the colored or shaded region in Fig. \ref{fig2}).
\par
\begin{figure*}[ht]
\centering
\includegraphics[height=2.5in,width=6.5in]{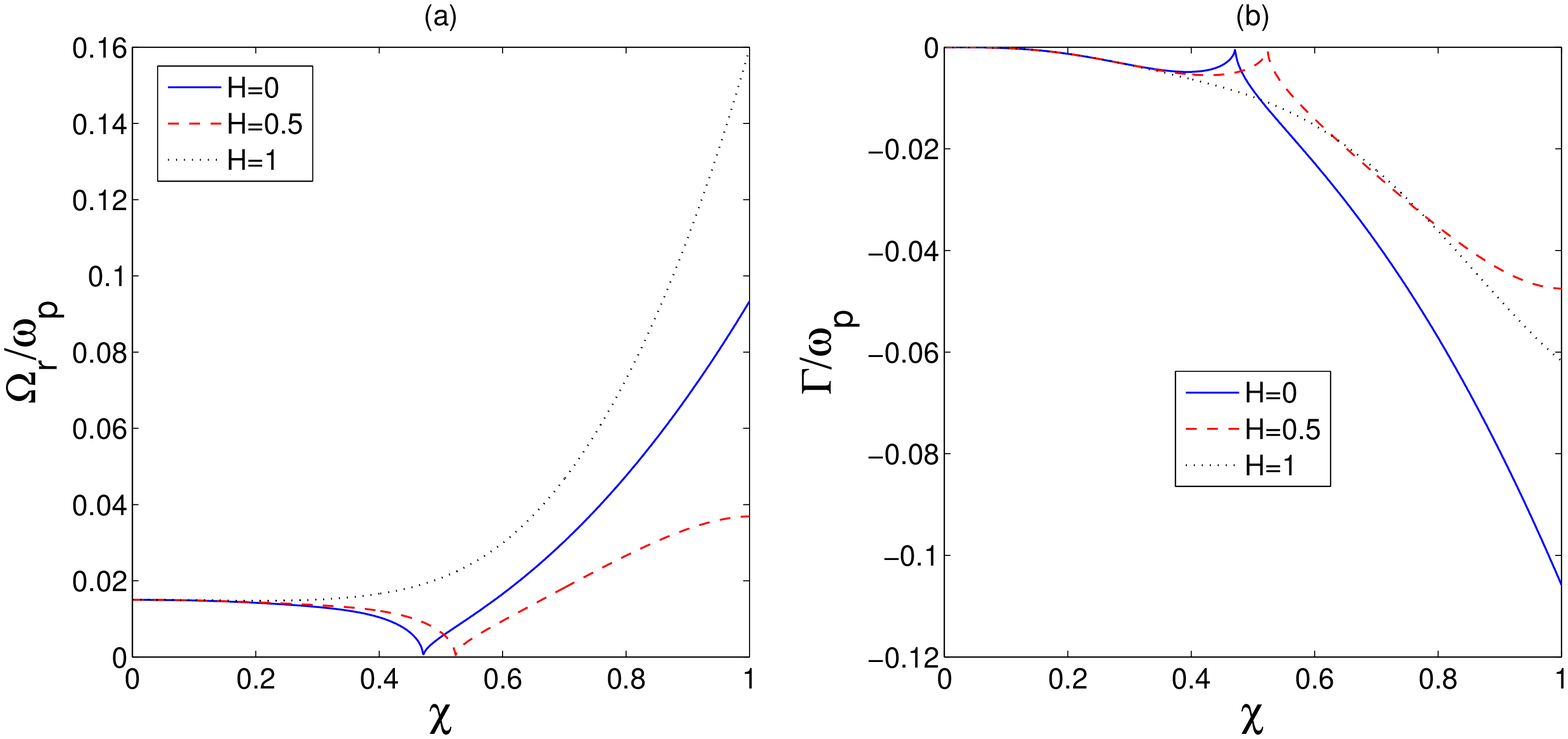}
\caption{ (Color online) The nondimensional frequency shift $\Omega_r/\omega_p$ [Eq. \eqref{Omega-real}, panel (a)]  and the energy transfer rate $\Gamma/\omega_p$ [Eq. \eqref{Omega-img},  panel (b)]  are plotted against the nondimensional carrier wave number $\chi\equiv k/k_d$ for different values of $H$ as in the legends and for a  fixed $\rho_0=K=0.1$.   }
\label{fig3}
\end{figure*}

 In a general manner, we also analyze the properties of $\Omega_r$ and $\Gamma$. The results are shown in  Fig. \ref{fig3}. The panel (a)  shows the   plot  of the frequency shift $\Omega_r$ and the panel (b) that for  the energy transfer rate $\Gamma$   against the nondimensional wave number $\chi$. With reference to the results in Fig. \ref{fig2}, where for some values of $\chi$ and $H$, the group velocity dispersion $(P)$ turns over going to zero and then to negative values, it is shown from Fig. \ref{fig3} that both $\Omega_r$ and $\Gamma$ have cutoffs exactly at the same $\chi$ where $P$ vanishes, and as $P$  becomes positive,   their absolute values increase with increasing values of  $\chi$. In fact, there are two regions of $\chi$, namely $0\lesssim\chi\lesssim\chi_0$ and $\chi_0\lesssim\chi\lesssim1$, where   $\chi_0$ is the  value of $\chi$ at which $P$ vanishes. In the former where $PQ\lesssim0$, the effect of $H~(\lesssim0.6)$   is to increase  both  the frequency shift $\Omega_r$ and the energy transfer rate $|\Gamma|$  having cutoffs at   higher values of $\chi$. However, in the other region where  $PQ>0$, the values of $\Omega_r$ and $|\Gamma|$ are seen to increase with $\chi$, but to decrease with $H$. Interestingly, for values of $H$ in the region $0.6<H\lesssim1$,  the behaviors of both $\Omega_r$ and $\Gamma$ are significantly changed, i.e., their absolute values start increasing with $\chi$ without any cutoffs (see the dotted lines in Fig. \ref{fig3}).       Furthermore, the maximum values (absolute) of both $\Omega_r$ and $\Gamma$ are achieved in $\chi_0\lesssim\chi\lesssim1$ with $H$ where  $PQ>0$.
\section{Nonlinear Landau damping of solitary wave solution}\label{sec-soliton-sol}
Following Ref. \onlinecite{chatterjee2015}, some approximate soliton solutions of the NLS equation \eqref{nls} with a small effect of the nonlinear Landau damping $(\propto R)$ can be obtained. 

When $PQ>0$, an approximate solitary wave solution of Eq. \eqref{nls}  is given by \cite{chatterjee2015}
\begin{equation}
\phi(\xi,\tau)=\sqrt{\phi_{0}(\xi,0)}\left(1-i\frac{\tau}{\tau_0}\right)^{-1/2}\text{sech}~{z}\exp(i\theta), \label{sol1-approx-nls}
\end{equation}

where  $z=(\xi-v_0\tau)/L$, $\theta= \left[v_0\xi+\left(\Omega_0-{v_0^2}/{2}\right)\tau\right]/2P$, with $v_0,~L,~\Omega_0,~\theta_1$ being constants, and $\tau_0$ is given by  
\begin{equation}
\begin{split}
\tau_0^{-1}=&\frac{\sqrt{2}R\phi_{0}(\xi,0)}{{\pi}^{3/2}\theta_1}\left[\frac{\cosh{(\pi\theta_1)}-1}{ \sinh{\left({\pi\theta_1}/{2}\right)}}\right]  {\cal{P}}  \int_{-\infty}^{\infty}\int_{-\infty}^{\infty}\\
&\times\left(\frac{\text{sech}^2 z'}{z-z'}\right) \text{sech}^2z \exp(i\theta_1z)dz dz'. 
\end{split} 
\end{equation}
 
On the other hand, for $PQ<0$, an approximate solitary wave solution of Eq. \eqref{nls}   is given  by \cite{chatterjee2015}
\begin{equation}
\phi={\phi_{0}(\xi,0)}\left(1-i\frac{\tau}{\tau_0}\right)^{-1/2}\text{tanh}{z}\exp(i\theta), \label{sol2-approx-nls}
\end{equation}
where $\theta$ and $\tau_0$ are given by 
\begin{equation}
\theta=\frac{1}{2P}\left[v_0\xi+ \left(2PQ\phi_0^2(\xi,0)-\frac{v_0^2}{2}\right)\tau\right], \label{theta-PQ<0}
\end{equation} 
\begin{equation}
\begin{split}
\tau_0^{-1}=&\left(\frac{2}{\pi}\right)^{3/2}\frac{R\phi_{0}^2(\xi,0)\left[1-\cosh(\pi\theta_2)\right]}{\delta(\tau)(1-\cosh\pi\theta_2)+\theta_2 \sinh\left(\frac{\pi\theta_2}{2}\right)}\\
&\times{\cal{P}} \int_{-\infty}^{\infty} \int_{-\infty}^{\infty} 
 \left(\frac{\tanh^2 z'}{z-z'}\right) \tanh^2z \exp(i\theta_2z)dz dz',
\end{split}
\end{equation}
with $\delta(\tau)$ denoting the Dirac delta function and $\theta_2$ a real constant.

\begin{figure}[ht]
\centering
\includegraphics[height=2.0in,width=3.5in]{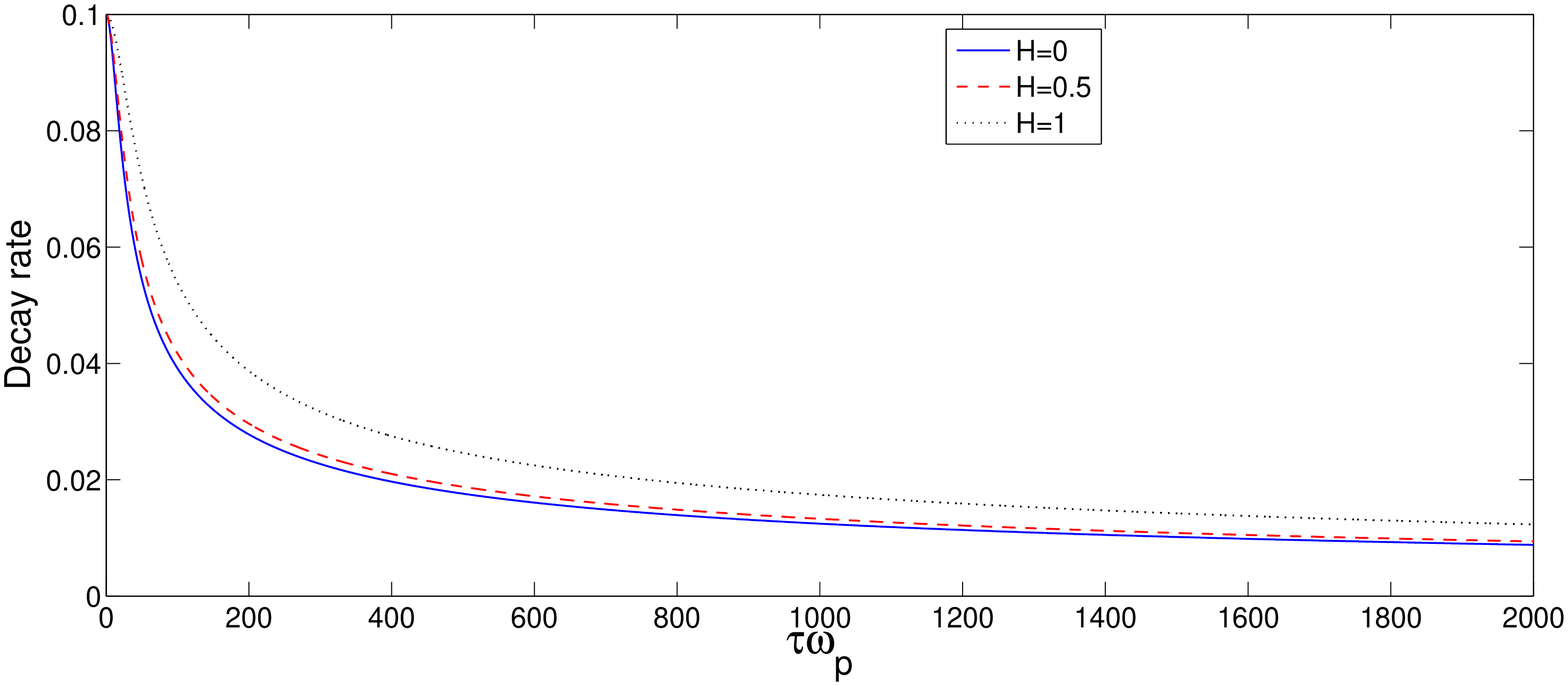}
\caption{The absolute value of the decay rate $| \left(1-i\tau/\tau_0\right)^{-1/2}|$ is shown against the nondimensional time $\tau\omega_p$ for different values of $H$ as in the legend. It is seen that  the decay of the wave amplitude is faster   the larger is the value of $H$. }
\label{fig4}
\end{figure}

A qualitative plot of the decay rate $| \left(1-i\tau/\tau_0\right)^{-1/2}|$ in Fig. \ref{fig4} shows that in both the cases of $PQ>0$ and $PQ<0$, the solitary wave amplitude decays with time due to the resonant particles having group velocity of the wave envelope, and the rate is relatively reduced compared to  its classical value by the effects of the quantum particle dispersion in EP plasmas. 
\section{Conclusion}
We have studied the influence of the effects of nonlinear wave-particle resonance   on the modulational instability and  nonlinear evolution of electrostatic wave envelopes in a collisionless quantum electron-positron-pair plasma. Starting from the Wigner-Moyal equation coupled to the Poisson equation, and using the multiple  scale technique, we have derived a NLS equation which governs the dynamics of wave envelopes in EP plasmas. In the weak quantum regime, it is shown that    the wave-particle resonance  occurs with the group velocity of the wave envelope, just similar to the   classical theory \cite{ichikawa1974}.  Also,  the coefficients of the NLS equation together with the nonlocal term, which appears due to the  resonance, are shown to be significantly modified by the quantum correction. Assuming the equilibrium distributions of electrons and positrons to be the Maxwellian and   the propagation of long-wavelength oscillations, explicit expressions for the     dispersion relation, the Landau damping rate, the coefficients of the NLS equation, as well as the frequency shift $(\Omega_r)$ and the energy transfer rate $(\Gamma)$  are obtained for modulated wave packets. 

 It is found that, by the effect of the quantum parameter $H$ (the ratio of plasmon energy to the thermal energy densities),     the Landau damping rate is greatly reduced compared to its classical value. Furthermore, the group velocity dispersion can turn over from negative to positive   values by the   effects of  $H$. This is in  consequence with the fact that both the frequency shift $\Omega_r$ and the energy transfer rate  $\Gamma$ vanish at a critical value of the carrier wave number $\chi\equiv k/k_d$. It is also found that in the region of $\chi$ where $PQ<0$, both the $\Omega_r$ and $|\Gamma|$ increase with increasing values of $H\lesssim0.6$ having cutoffs at higher values of $\chi$, however, opposite behaviors are noticed in the other region where $PQ>0$. Here,   both $\Omega_r$ and $|\Gamma|$ increase   with increasing values of $\chi$ and but decrease with $H$. However, for values of $H$ in  $0.6\lesssim H\lesssim1$, both $\Omega_r$ and $|\Gamma|$ gradually increase with $\chi$ having no cutoffs.  
Some approximate solitary wave solutions of the NLS equation are also obtained with a small effect of the nonlinear Landau damping. It is found that the quantum parameter $H$ slows down the  decay rate of the solitary wave amplitude in  EP plasmas. 

We mention that the inclusion of the other quantum effects, e.g., the exchange/correlation force, in the present theory  can be equally important \cite{zamanian2013,ekman2015} as its relative contribution in the wave dispersions and nonlinearities roughly scale as $~c_0H^2$, where $c_0$ is some constant factor. It has been found that for  high-frequency Langmuir waves, the contribution of the exchange effects in the linear dispersion relation is relatively small with $c_0\sim1/4\pi$ \cite{zamanian2015}. However, in the nonlinear regime, even though its contribution will still scale as $~c_0H^2$, the factor $c_0$ may or may not be smaller than the unity. Furthermore, dealing with exchange effects in the nonlinear regime requires a    challenging task both from computational and numerical points of view, and thus remains a project for future study. Other important aspects, which we have limited in this work, could be the influence of the particle dispersion in the strong quantum regime where $\hbar k/mv_t$ is not small. 

 Further to mention that in the limit of $T>T_F$, where $T_F$ is the Fermi temperature, while   the  Pauli blocking mechanism reduces, the collisional effects between the charged particles may increase.  In principle,   one can add a collisional operator (e.g., Coulomb collision) to the semiclassical Vlasov equation \eqref{wigner-eq} similar to the Vlasov-Poisson system, however, it may not be easy to include so in the Wigner or Wigner-Moyal equation \eqref{wigner-eq-moyal}. Nevertheless, inclusion of such collisional term contributes to the NLS equation \eqref{nls} as a linear damping-like term which may be comparable to or smaller than  the term $\propto S\sim\gamma_L/\epsilon^2$ associated with the linear Landau damping rate $\gamma_L\sim\epsilon^3$.      In the present analysis, the nonlocal nonlinearity (which violates the conservation laws associated with the ordinary NLS equation), due to the  nonlinear resonance (with the group velocity of the wave),   dominates the dissipative mechanism over those produced by the collisional and  linear  Landau damping effects. Furthermore, we have considered the time scale $(\sim\epsilon^{-2})$ which is much longer than the bounce period of trapped particles $\tau_{trap}\sim\epsilon^{-1/2}~(\sim \nu $, the Coulomb collision rate).   So, to some extent, it is safe to neglect the collision effects  in the present model. 
\par
 To conclude,  our results will  be equally well-applicable to other pair plasmas or electron-ion plasmas where ions are treated classical, as one can simply recover the results by the adjustment of some  factors in the relevant expressions. 
\acknowledgments{The authors sincerely thank Professor Gert Brodin of Ume{\aa} University, Sweden, for  his valuable comments and suggestions which improved the manuscript in its present form.  This work was supported by UGC-SAP (DRS, Phase III) with  Sanction  order No.  F.510/3/DRS-III/2015(SAPI),  and UGC-MRP with F. No. 43-539/2014 (SR) and FD Diary No. 3668.}
\appendix
\section{Expressions for $A_1,~B,~\Theta,~\Phi$ etc.}\label{appendix-a}
Here, we give the expressions for $A_1,~B,~\Theta,~\Phi$ etc. appearing in Eq. \eqref{gamma}: 
\begin{equation}
\begin{split}
A_1=& 4\pi \sum_\alpha \frac{e_\alpha^2}{m_\alpha}\int \frac{1}{\omega-kv}\frac{\partial}{\partial v}\left(\frac{G_\alpha-k^2 H_\alpha}{\omega-kv} \right)dv\\
& +k^24\pi \sum_\alpha \frac{e_\alpha^3}{m^2_\alpha}\int \frac{1}{\omega-kv}\frac{\partial}{\partial v}\left(\frac{3 H_\alpha}{\omega-kv} \right)dv\\
&+k^2 4\pi \sum_\alpha \frac{e_\alpha^2\hbar^2}{24 m^3_\alpha}\int \frac{1}{\omega-kv}\frac{\partial^3}{\partial v^3}\left(\frac{G_\alpha-k^2 H_\alpha}{\omega-kv} \right)dv\\
&+k^4 4\pi \sum_\alpha \frac{e_\alpha^2\hbar^2}{24m^3_\alpha}\int \frac{1}{\omega-kv}\frac{\partial^3}{\partial v^3}\left(\frac{3 H_\alpha}{\omega-kv} \right)dv, \label{A1}
\end{split}
\end{equation}
\begin{equation}
\begin{split}
B=&4\pi \sum_\alpha \frac{e^3_\alpha}{m^2_\alpha} \int \frac{1}{\omega-kv} \frac{\partial}{\partial v} \left[ \frac{1}{\omega-kv} \frac{\partial}{\partial v}\left( \frac{G_\alpha-k^2H_\alpha}{\omega-kv}\right) \right]\\ &\times dv+k^2 4\pi \sum_\alpha \frac{e^3_\alpha\hbar^2}{24 m^4_\alpha} \int \frac{1}{\omega-kv} \frac{\partial^3}{\partial v^3} \left[ \frac{1}{\omega-kv} \right.\\ 
 &\left.\times\frac{\partial}{\partial v}\left( \frac{G_\alpha-k^2H_\alpha}{\omega-kv}\right) \right] dv
 +k^2 4\pi \sum_\alpha \frac{e^3_\alpha\hbar^2}{24 m^4_\alpha} \\
 &\times\int \frac{1}{\omega-kv} \frac{\partial}{\partial v} 
  \left[ \frac{1}{\omega-kv} \frac{\partial^3}{\partial v^3}\left( \frac{G_\alpha-k^2H_\alpha}{\omega-kv}\right) \right] dv \\
&+k^4 4\pi \sum_\alpha \left( \frac{e_\alpha\hbar^2}{24 m^3_\alpha}\right)^2 e_\alpha \int \frac{1}{\omega-kv} \\
&\times\frac{\partial^3}{\partial v^3} \left[ \frac{1}{\omega-kv}\frac{\partial^3}{\partial v}\left( \frac{G_\alpha-k^2H_\alpha}{\omega-kv}\right) \right] dv, \label{B}
\end{split}
\end{equation}
\begin{equation}
\begin{split}
\Theta(k,\omega)=&k^3\left[ \frac{\Delta}{\Delta^2+\Gamma^2}(WW_3-UU_3)\right.\\
&\left.+ \frac{2\Gamma}{\Delta^2+\Gamma^2}WUW_3U_3+C_3\right], \label{Theta-k-omega}
\end{split}
\end{equation}
\begin{equation}
\begin{split}
\Phi(k,\omega)=&k^3\left[\frac{\Gamma}{\Delta^2+\Gamma^2}(WW_3-UU_3)\right.\\
&\left.-  \frac{2\Delta}{\Delta^2+\Gamma^2}WUW_3U_3+D_3\right], \label{Phi-k-omega}
\end{split}
\end{equation}
\begin{equation}
\begin{split}
 &W_3= W_1+6k^4W_2,~~ U_3=U_1+6k^4U_2,\\
 & C_3= C_1+6k^4C_2,~~ D_3=D_1+6k^4D_2,
 \end{split}
\end{equation}
\begin{equation}
\begin{split}
&W(\lambda)=4\pi \sum \frac{e^2_\alpha}{m_\alpha}\int \frac{I_\alpha(v)}{v-\lambda}dv,\\
&U(\lambda)=4\pi^2 \sum_\alpha \frac{e_\alpha ^2}{m_\alpha} I_\alpha (\lambda), \label{W-k-omega-lamb}
\end{split}
\end{equation}
\begin{equation}
\begin{split}
&W_1(k,\omega;\lambda)=4\pi \sum \frac{e^2_\alpha}{m_\alpha}\int \frac{1}{(\omega-kv)^2}\frac{G_\alpha}{v-\lambda}dv,\\
&U_1(k,\omega; \lambda)=4\pi^2 \sum_\alpha \frac{e_\alpha ^2}{m_\alpha} \frac{G_\alpha (\lambda)}{(\omega-k\lambda)^2}, \label{W1-k-omega-lamb}
\end{split}
\end{equation}
\begin{equation}
\begin{split}
&W_2(k,\omega;\lambda)=4\pi \sum \frac{e^2_\alpha\hbar^2}{24m^3_\alpha}\int \frac{1}{(\omega-kv)^4}\frac{G_\alpha}{v-\lambda}dv,\\
&U_2(k,\omega; \lambda)=4\pi^2 \sum_\alpha \frac{e_\alpha ^2\hbar^2}{24m^3_\alpha} \frac{G_\alpha (\lambda)}{(\omega-k\lambda)^4}, \label{W2-k-omega-lamb}
\end{split}
\end{equation}
\begin{equation}
\begin{split}
&C_1(k,\omega; \lambda)=-4\pi\sum_\alpha \frac{e_\alpha^3}{m_\alpha^2}\int \frac{1}{(\omega-kv)^2} \frac{I_\alpha(v)}{v-\lambda}dv,\\
&D_1(k,\omega; \lambda)=4\pi^2\sum_\alpha \frac{e_\alpha^3}{m_\alpha^2}\frac{I_\alpha(\lambda)}{(\omega-k\lambda)^2}, \label{C1-k-omega}
\end{split}
\end{equation}
\begin{equation}
\begin{split}
&C_2(k,\omega; \lambda)=-4\pi\sum_\alpha \frac{e_\alpha^3\hbar^2}{24m_\alpha^4}\int \frac{1}{(\omega-kv)^4} \frac{I_\alpha(v)}{v-\lambda}dv,\\
&D_2(k,\omega; \lambda)=4\pi^2\sum_\alpha \frac{e_\alpha^3\hbar^2}{24m_\alpha^4}\frac{I_\alpha(\lambda)}{(\omega-k\lambda)^4}, \label{C2-k-omega}
\end{split}
\end{equation}
 \begin{equation}
 \begin{split}
&\Delta^{(c)}(K,\Omega)=\Delta(\lambda)+i\Gamma(\lambda)\frac{K}{|K|},\\
&\Delta(\lambda)=4\pi \sum_\alpha e_\alpha \int \frac{G_\alpha(v)}{v-\lambda}dv,~~\Gamma(\lambda)=4\pi^2\sum_\alpha e_\alpha G_\alpha(\lambda). \label{Delta-c-final}
\end{split}
\end{equation}
 Note that the functions $U_1,~U_2,~D_1,~D_2$ and $\Gamma$ represent contributions of the resonant particles having the group velocity of the wave envelope in EP plasmas.
  \section{Various expressions to derive P} \label{appendix-b}
  We calculate the various terms (in the limit $\chi^2\ll1$) which appear in the coefficients of the  NLS equation \eqref{nls} for the Maxwellian distribution \eqref{maxwellian-distribution}   as follows:
\begin{equation}
\alpha= -\frac{2k}{\omega_p}\left[ 1+ \frac{3}{2}\chi^2 +\frac{H^2}{8}\chi^4\right], \label{alpha-coeff}
\end{equation}
\begin{equation}
\beta=-\frac{3}{k_d}\chi \left[ 1- \left( 6-\frac{H^2}{2}\right) \chi^2 +\frac{23}{2} H^2\chi^4 \right], \label{beta-coeff}
\end{equation}
\begin{equation}
\begin{split}
&A=A_1=0,~~W=W_1=W_2=0,\\
&W_3=W_1+6k^4W_2=0,\\
& U=U_1=U_2=0;~~U_3=U_1+6k^4U_2=0, \label{AWU-coeff}
\end{split}
\end{equation}
\begin{equation}
B= -15\left( \frac{e}{m}\right) ^2 \frac{k^3}{{\omega_p}^4}\left[ 1+ \left( 12+\frac{H^2}{12}\right)\chi^2 - \frac{ H^2}{6} \chi^4\right] ,\label{B-coeff}
\end{equation}
\begin{equation}
\Delta= -k_d^2 [1-9\chi^2 -3H^2\chi^4],\label{Delta-coeff}
\end{equation}
\begin{equation}
\Gamma=-\left(\frac{9\pi}{2} \right)^{1/2}k_d^2\chi \left[ 1-\left(6-\frac{H^2}{6} \right)\chi^2 -\frac{H^2}{8}  \chi^4  \right]  , \label{Gamma-coeff}
\end{equation}
\begin{equation}
C_1=\left(  \frac{e}{k_BT}\right) ^2 \frac{1}{k_d^2} \left[1+\left(27-\frac{H^2}{4}\right) \chi^2 -4H^2\chi^4  \right] , \label{C1-coeff}
\end{equation}
\begin{equation}
C_2=-\frac{H^2}{24k_d^6}\left(\frac{e}{k_BT} \right)^2 \left(1-21\chi^2\right), \label{C2-coeff} 
\end{equation}
\begin{equation}
\begin{split}
D_1=&-\left(  \frac{e}{k_BT}\right) ^2 \frac{1}{k_d^2} \left( \frac{9\pi}{2}\right) ^{1/2} \\
&\times \chi \left[ 1- \left(\frac{9}{2}+\frac{13H^2}{24} \right) \chi^2-\frac{57}{16} H^2 \chi^4\right] . \label{D1-coeff}
\end{split}
\end{equation}
\begin{equation}
D_2=\frac{H^2}{24k_d^6} \left(  \frac{e}{k_BT}\right) ^2   \left( \frac{9\pi}{2}\right) ^{1/2}  \chi \left( 1+13\chi^2\right), \label{D2-coeff}
\end{equation}
\begin{equation}
\begin{split}
C_3&=C_1+6k^4C_2\\
&=\left(  \frac{e}{k_BT}\right) ^2 \frac{1}{k_d^2}\left[1+\left(27-\frac{H^2}{4}\right) \chi^2- \frac{7H^2}{2}\chi^4  \right], \label{C3-coeff}
\end{split}
\end{equation}
\begin{equation}
\begin{split}
D_3&=D_1+6k^4D_2\\
&=-\left(  \frac{e}{k_BT}\right) ^2 \frac{1}{k_d^2} \left( \frac{9\pi}{2}\right) ^{1/2} \\
&\times \chi \left[ 1- \left(9+\frac{13H^2}{24} \right)  \chi^2- \frac{53H^2}{16} \chi^4 \right]. \label{D3-coeff}
\end{split}
\end{equation}


\begin{thebibliography}{50}
\bibitem{landau1946} L. Landau,  Zh. Eksp. Teor. Fiz. \textbf{16}, 574 (1946) [J. Phys. USSR
\textbf{10}, 25 (1946)].
\bibitem{malmberg1964} J. H. Malmberg and C. B. Wharton, Phys. Rev. Lett. \textbf{13}, 184 (1964).
\bibitem{rightley2016} S. Rightley and D. Uzdensky, Phys. Plasmas \textbf{23}, 030702 (2016).
\bibitem{chatterjee2015} D. Chatterjee and A. P. Misra, Phys.Rev.E \textbf{92}, 063110 (2015).
\bibitem{villani2014} C. Villani, Phys. Plasmas \textbf{21}, 030901 (2014).
\bibitem{zheng2013} J. Zheng and H. Quin, Phys. Plasmas \textbf{20}, 092114 (2013).
\bibitem{valentini2007} F. Valentini and R. D'Agosta, Phys. Plasmas \textbf{14}, 092111 (2007).
\bibitem{mukherjee2014} A. Mukherjee, A. Bose, and M. S. Janaki, Phys. Plasmas \textbf{21}, 072303 (2014).
\bibitem{brodin2015} G. Brodin, J. Zamanian, and J. T. Mendonca, Phys. Scr. \textbf{90}, 068020 (2015).
\bibitem{ichikawa1974} Y. H. Ichikawa, Suppl. Prog. Theor. Phys. \textbf{55}, 212 (1974).
\bibitem{ichikawa1972} Y. H. Ichikawa, T. Imamura, and T. Taniuti, J. Phys. Soci. Jpn \textbf{33}, 189 (1972).
\bibitem{sarri2015} G. Sarri,	K. Poder,	J. M. Cole,	W. Schumaker,	A. Di Piazza,	B. Reville,	T. Dzelzainis,	D. Doria, L. A. Gizzi,	G. Grittani, 	S. Kar,	C. H. Keitel,	K. Krushelnick,	S. Kuschel,	S. P. D. Mangles,	Z. Najmudin,	 N. Shukla,	L. O. Silva, 	D. Symes,	A. G. R. Thomas,	 M. Vargas,	J. Vieira,	and M. Zepf, Nat. Commun. {\bf 6}, 6747 (2015).
\bibitem{manfredi2001} G. Manfredi and F. Haas, Phys. Rev. B \textbf{64}, 075316 (2001); G. Manfredi, Fields Inst. Commun. \textbf{46}, 263 (2005).
\bibitem{chen1984} F. F. Chen, \textit{Plasma Physics and Controlled Fusion} (Second edition, Vol. 1, Plenum Press, New York, 1984), pp. 291.
\bibitem{zamanian2013} J. Zamanian, G. Brodin, and M. Marklund, Phys. Rev. E   \textbf{88}, 063105 (2013). 
\bibitem{ekman2015} R. Ekman, J. Jamanian, and G. Brodin, Phys. Rev. E \textbf{92}, 013104 (2015).
\bibitem{zamanian2015} J. Zamanian, G. Brodin, and M. Marklund, European Phys. J. D \textbf{69}, 25 (2015). 




\end{thebibliography}
\end{document}